\documentclass[useAMS,usenatbib]{mn2e}
\usepackage{graphicx}
\usepackage{epstopdf}
\usepackage[colorlinks = true, linkcolor = blue, urlcolor  = blue, citecolor = blue, anchorcolor = blue]{hyperref}
\newcommand\fnurl[2]{\href{#2}{#1}\footnote{\url{#2}}}

\title[Spatio-temporal distribution of precursor flares]{Statistical study of spatio-temporal distribution of precursor solar flares associated with major flares}

\author[N. Gyenge et al.]{N. Gyenge$^{1,2}$\thanks{E-mail,
n.g.gyenge@sheffield.ac.uk}, I. Ballai$^{2}$, T. Baranyi$^{1}$
\\
$^{1}$Debrecen Heliophysical Observatory (DHO), Konkoly Observatory, Research Centre for Astronomy and Earth Sciences\\Hungarian Academy of Sciences, 4010 Debrecen, P.O. Box 30, Hungary\\
$^{2}$Solar Physics and Space Plasmas Research Centre (SP2RC), University of Sheffield,\\
Hounsfield Road, Hicks Building, Sheffield S3 7RH, UK
}

\begin{document}
\date{}
\pagerange{\pageref{firstpage}--\pageref{lastpage}}
\pubyear{}
\maketitle
\label{firstpage}

\begin{abstract}

The aim of the present investigation is to study the spatio-temporal distribution of precursor flares during the 24-hour interval preceding M- and X-class major flares and the evolution of follower flares. Information on associated (precursor and follower) flares is provided by Reuven Ramaty High Energy Solar Spectroscopic Imager (RHESSI). Flare List, while the major flares are observed by the Geostationary Operational Environmental Satellite (GOES) system satellites between 2002 and 2014. There are distinct evolutionary differences between the spatio-temporal distributions of associated flares in about one day period depending on the type of the main flare. The spatial distribution was characterised by the normalised frequency distribution of the quantity $\delta$ (the distance between the major flare and its precursor flare normalised by the sunspot group diameter) in four 6-hour time intervals before the major event. The precursors of X-class flares have a double-peaked spatial distribution for more than half a day prior to the major flare, but it changes to a lognormal-like distribution roughly 6 hours prior to the event. The precursors of M-class flares show lognormal-like distribution in each 6-hour subinterval. The most frequent sites of the precursors in the active region are within a distance of about 0.1 diameter of sunspot group from the site of the major flare in each case. Our investigation shows that the build-up of energy is more effective than the release of energy because of precursors.

\end{abstract}


\section{Introduction}

Solar flares are one of the most powerful energetic events in the solar atmosphere (for a comprehensive review see, e.g.  \citealt{Benz}, \citealt{Fletcher}). Given their role in the energy balance of the solar corona and their role played in driving space weather, many studies investigated the energy build-up and initiation of flares, concentrating on the events preceding the onset of a flare. 

The early studies of X-ray class flares have shown that the solar conditions leading to flare activity over a period of one day change sufficiently slowly that the probability of occurrence of a flare event does not show significant daily change \citep{Neidig}. The comprehensive study by \cite{Tappin} has shown that the vast majority of hard X-ray (HXR) flares are preceded by one or more soft X-ray bursts called \textit{precursors} some 10 to 60 minutes before the major flares. In some cases, the precursor can be classified as a flare in its own right but, by definition, a precursor cannot be stronger than its associated flare. The properties of precursors range from significant flares right down to microflares, but they are usually several orders of magnitude weaker than the main flare. It is also shown that precursor-like bursts can be observed in the absence of subsequent energetic flare, although they occur at a lower rate than in the hour prior to a major flare. Tappin's study mentions that the information on precursors seem to imply some process whereby either the energy releases progressively destabilise the magnetic structure of the active region, or else they occur into a region where the magnetic structure is becoming progressively less stable.

During the last decades, a large number of studies contributed to revealing of the characteristics and nature of the processes that may influence or determine the flare occurrence on various time-scales, by building up of energy and triggering of flares. The most important of such known slowly changing conditions and slow processes (magnetic complexity, new flux emergence, flux cancellation, shear motions, sunspot rotation, and magnetic helicity injection) and some rapid changes of magnetic fields have been reviewed recently by \cite{WangLiu}. The studies that investigated the major flares and the evolution and characteristics of their precursors over longer or shorter periods have applied various methods. In order to lay the foundation and motivation of our research, we list the key methodological differences of the studies found in the literature dealing with these questions.

There are statistical studies concentrating on the changes occurring in an active region (AR) in the 1-hr time interval before a major flare. These studies usually contain about 20-100 major flares of various types (C,M, and X), however, the short time interval used in these studies prevent them from diagnosing confidently the temporal and spatial evolution of ARs (for studies of pre-flaring covering time periods spanning from minutes to max. one hour, see e.g. \cite{Farnik96}, \cite{Farnik98}).     

Several case studies describe one or a few major flares with a few precursors. They can help in revealing the physical background of the investigated particular cases but they are not suitable for determining how the given processes cause or contribute to the build-up or onset of the main flare. For example, \cite{Liu} found that four flares and two coronal mass ejections (CMEs) occurred with a causal relationship in an active region during a 1.5-hour period. \cite{Joshi} studied a series of events consisting of four small flares followed by a major flare within two hours. They found that the small flares were manifestations of localised magnetic reconnection at different evolutionary stages of the filament, and they played a key role in destabilising the filament leading to a major eruption. \cite{Chifor} have concluded that the X-ray precursors observed between 2 and 50 minutes before the start of the filament fast-rise and flare impulsive phase may be connected to a process called tether-cutting mechanism leading to flare and filament eruption. \cite{Sterling05} reported that newly emerged flux was observed near the location of the initial X-ray brightenings about 2 days prior to onset of the main flare and filament eruption. They concluded that this recently emerged flux could have been a catalyst for initiating the tether-cutting reconnection. As it can be seen in the H${\alpha}$ flare event list of National Geophysical Data Center (NGDC), several small flares were observed in that region during these 2 days of flux emergence.

In addition, there are a few several statistical studies that used large datasets. For example, the waiting-time distribution of GOES flares (i.e. the distribution of times between flares) is analysed in a number of studies (e.g. \cite{W00}, \cite{WL02}), but these studies do not investigate the possible relationship between precursors and major flares.  \cite{Falconer} used the data of 1300 ARs to show that prior to major flaring event, these -in addition to reflecting the amount of free energy in the active region- reflect significant determinants of coming productivity of major eruptions.  However, these types of studies are based on the flaring rate, i.e. number of flares per day, thus their study cannot provide information on the changes in the 24-hour interval before major flares.

Some conditions leading to flares may change over a period shorter than 24 hours, but longer than 1 hour. It is reasonable to assume that the effect of these changes may also be detected in the precursor flaring activity during the 24-hour interval before the main flare. The literature on this topic is not too vast, the number of papers dealing with investigation for such an extended time interval is very low.

For example, \cite{Jakimiec} used a medium-sized statistical sample, concluded that the precursors are often placed in the vicinity of strong flares for such cases when the precursor precedes the strong flare in time interval shorter than 24 hours. \cite{Balazs} used a large sample of data containing about 18,500 flares observed by the RHESSI satellite. They have shown that there is a statistical relationship between temporal and spatial distances of succeeding flares. They also found a power law connection between these parameters within a period of about one day. Their data also support the assumption of lognormal distribution of spatial distances in the case of time difference smaller than about 2-3 hours. Although in this study the relationships between the precursor and main flares have not been investigated, the results may refer to precursor flares because the majority of the studied flares were small flares observed in the 6-12 keV channel. Thus, it seems reasonable to assume that their results may show the  characteristics of the precursor flares and the related processes ongoing in this period. The assumption of the existence of processes effective on this time-scale is supported by \cite{Korsos} who presented a few examples showing that the intensity of RHESSI flares  integrated in 4-hour or 12-hour bins starts increasing well before the energetic X-class flares. 

The investigation of changes of spatial and temporal characteristics of precursor flares during the 24-hr interval before major X- and M-type flares may provide information on such a segment of the observational phase space that has not been well-known so far. The location of precursors may indicate the region where the magnetic structure progressively changes by successive reconnection events. The temporal distribution of occurrence of associated flares may reveal the statistically characteristic time at which the destabilisation usually starts before major flares.

\section{Data processing and selection criteria}

\begin{figure*}
	\center
	\includegraphics[width=130mm]{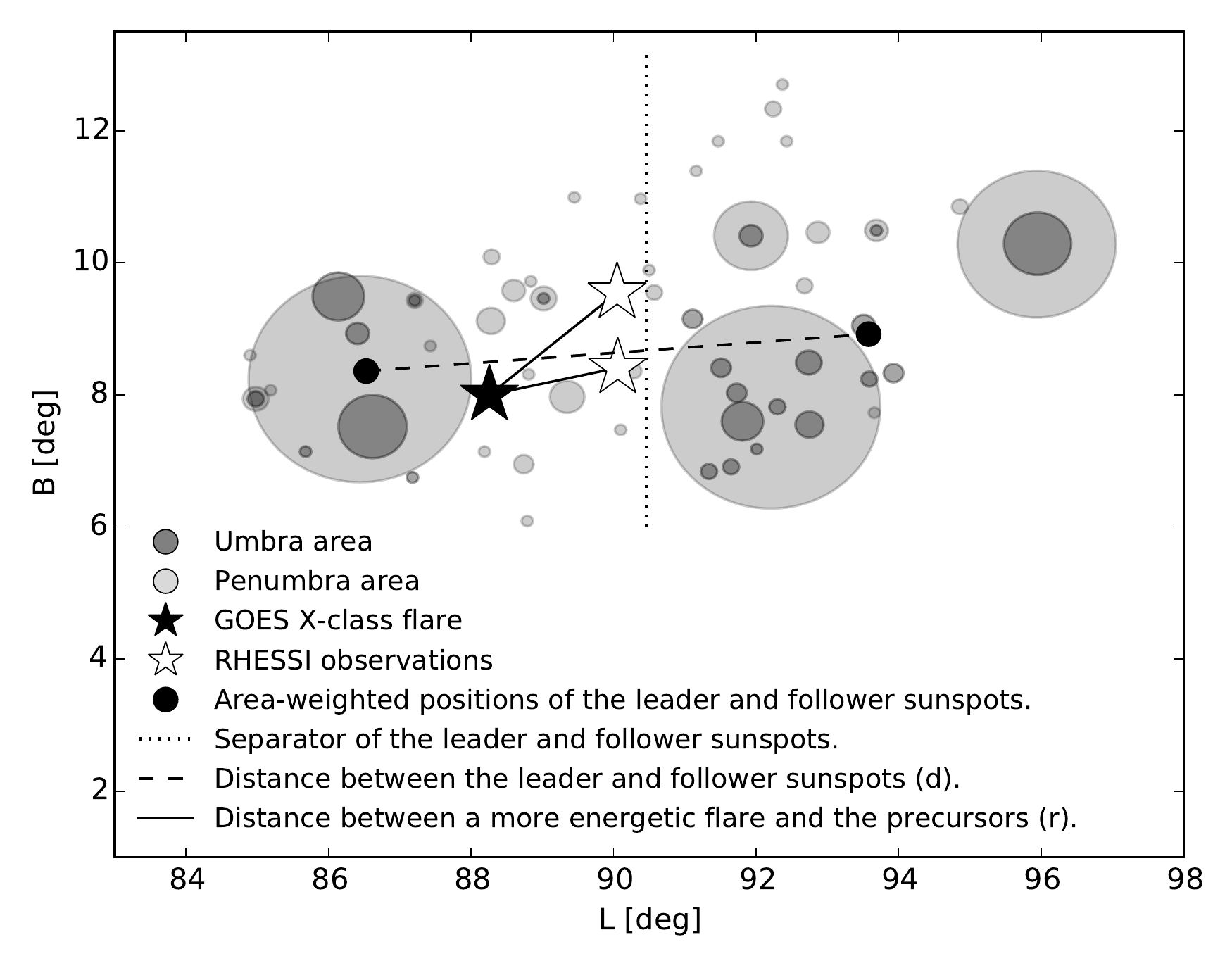}
	\caption{Schematic representation of a sample of data involved in the study. The figure shows the drawing of the active region NOAA 11166 (2011-03-09 17:44:58) which is reconstructed from the DPD catalogue. The umbrae and penumbrae of the sunspot group are distinguished by the different shade of grey colour. The area of a light grey circle corresponds to the related whole spot area measured in msh (millionths of solar hemisphere). The area of a dark grey circle is the same for umbra area.  The black circles represent the area-weighted centroids of the leading and following parts separated by the dotted line. The quantity $d$ represent the distance between the centroids. The two white stars denote the positions of two RHESSI precursor flares and the black star shows the position of the main flare recorded by GOES. The two solid line segments (quantified by the quantity $r$) represent the distances between the positions of GOES and RHESSI events.}
	\label{dpd}
\end{figure*}

Our study uses two X-ray flare databases. The major flares were collected using the dataset provided by the \fnurl{Geostationary Operational Environmental Satellite}{http://www.ngdc.noaa.gov/stp/satellite/goes/} (GOES). GOES flares are classified as A, B, C, M, or X-class, according to their peak flux (W m$^{-2}$) observed in the $0.1$ to $0.8$ nm wavelength range. We selected the M- and X-class flares corresponding to a flux in excess of $10^{-5}$ W m$^{-2}$ and $10^{-4}$ W m$^{-2}$ at Earth, respectively. The GOES flare lists are available at \fnurl{NGDC/NOAA}{ftp://ftp.ngdc.noaa.gov/STP/space-weather/solar-data/solar-features/solar-flares/x-rays/goes/}. In the GOES flare lists, the flares of C-class or larger are only listed (except for a few cases) and the position data or the the numbers of the associated active regions are often missing. Thus, the GOES flare lists are not suitable for studies of small flares but they are the only sources of the major (M- and X-class) flare events with information on their peak X-ray fluxes.  

The smaller and less energetic X-ray flares occurring before the onset of major GOES events have been examined by using the data provided by the \fnurl{RHESSI}{http://hesperia.gsfc.nasa.gov/hessidata/dbase/} satellite \citep{Lin}. Since its launch in 2002, RHESSI observed more than 100,000 events measured in the energy band from soft X-ray to gamma ray. The RHESSI flares are mostly small flares that correspond to GOES A-, B-, or C-class flares, with the most frequent type of flares being GOES class B but the more rare major flares are also observed. The RHESSI flare list always contains the position data (in arcsecond) of a flare if it was observed in the 6-12 keV or larger energy band but it does not contain information on the GOES class. Thus, the RHESSI Flare List is much suitable for statistical studies of spatial and temporal distribution of smaller flares than the GOES flare list.

Our statistics includes only those major events that were observed simultaneously by the GOES and RHESSI satellites, i.e. the main flare can be identified in both datasets. The selection criteria for major flares were the following: (1) the position of the candidate flare was published in the RHESSI flare list (we used the GOES flare list as a confirmation of the flare classification) and the associated active region was identified;  (2) there was no flare with a larger peak flux in the given active region within the 24-hour interval before the candidate, i.e. each flare that appears prior to the candidate can be reckoned as a precursor flare, and the candidate is the main flare; (3) there was no stronger follower flare in the given active region within the next 24 hours after the candidate, i.e. we exclude those situations when the candidate is considered as a precursor of a next larger flare; (4) in order to avoid the question whether an X-class flare can be a precursor of a stronger X-class flare, we de-selected those cases when an X-class flare was preceded by an another X-class event within a day interval (both belonging to the same active region). Less than $10 \%$ of the original data was omitted for this reason; (5) we omitted all those candidates that erupted on or near the solar limb ($\pm 70^\circ$ from the central meridian) in order to decrease the error of the position determination caused by projection effects. Each flare in the RHESSI flare list occurring in the 24-hour interval before and after the major flare and fulfilling the selection criteria is taken into account and considered as an associated flare of the main flare.

To derive spatial distances between the locations of flares, we converted the ($x,y$) positions of flares [arcsec] to latitude and longitude [deg] of the Carrington heliographic coordinate system. The method is described by \cite{Thompson}.The position of flares are in the  6 - 12 keV energy band. Therefore, we have information about the energetic events occurring at the top of magnetic loops, ensuring that we did not mix the top of the loops with the foot-point position. As a result, our sample that matches all the above selection criteria contains 49 X-class flares (associated with 1001 precursor and 345 follower flares) and 315 M-class events (associated with 3151 precursor and 1512 follower flares). 

The Debrecen Photoheliographic Data sunspot catalogue (DPD) \citep{DPD} has been used to estimate the morphological properties of the associated sunspot groups based on white-light full-disk observations. The catalogue provides information about the following properties of the every sunspot: date of observation, position of the spot, and umbra and whole sunspot area. The positions of the leading and following parts of sunspot groups have been taken from the \fnurl{DPD tilt angle database}{http://fenyi.solarobs.unideb.hu/test/tiltangle/dpd/}. In this dataset, the area-weighted heliographic longitude of the sunspots (the longitude of the centroid of the whole group) separates the leading and following parts and the corresponding area-weighted positions have been determined for both parts of the sunspot groups \citep{Baranyi}. We used those data that are derived by using the whole spot area as weight. The temporal resolution of DPD is one observation for each day. All the sunspots are considered for the date of the main flare. 

\section{Spatio-temporal evolution of associated flares}
\subsection {Spatial distribution of precursor flares}

\begin{figure}
	\center	
	\includegraphics[width=85mm]{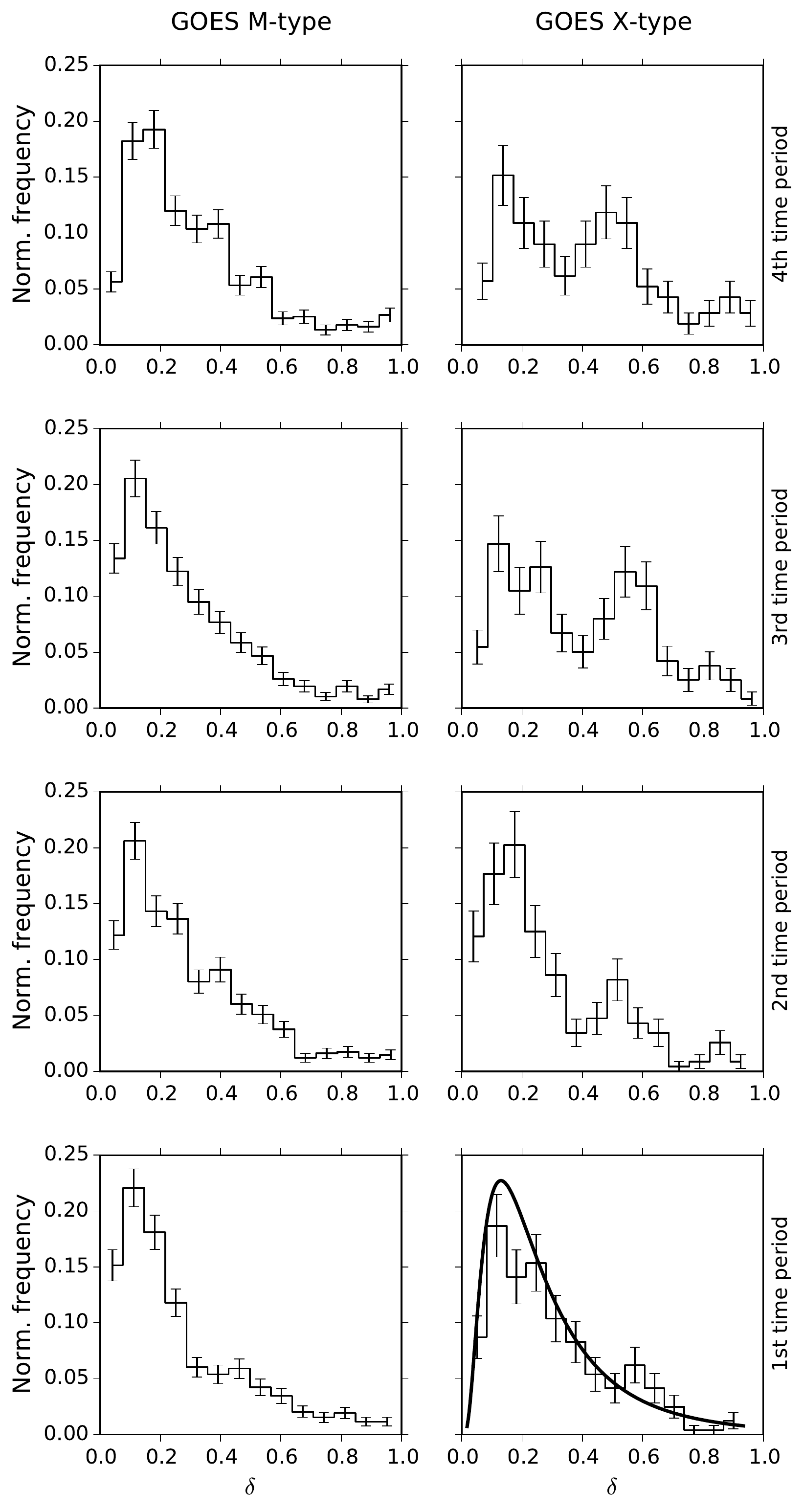}
	\caption{The normalised frequency distribution of the quantity $\delta$ in the 24-hour interval prior to the major GOES  M- (left hand side panels) and X-class (right hand side panels) flares. The different rows show the results derived in successive time intervals in bins of 6 hours. The 1st period (0-6 hours) is the closest in time to the major flare. The error bars are derived by modelling the number of events observed in each bin as a Poisson variable.}
	\label{NFD}
\end{figure}

We carry out statistical analysis for the selected M- and X-class flares in order to establish the spatial distributions of their precursors and the possible temporal variation of their spatial distribution.

We define the quantity $r$ as a distance between the main flare detected by GOES and its RHESSI precursor. The quantity $2d$ is the estimated diameter of the whole sunspot group where $d$ is the distance between the centroids of the leading and following parts of sunspot group.  Figure \ref{dpd} represents a typical example for an active region where these quantities are defined. To carry out our statistical analysis, we define the quantity $\delta$ as,

\begin{equation}
	\delta=r/2d.
\end{equation}

The quantity $\delta$ is a relative distance that estimates the distance of the precursor and the main flare relative to the size of the sunspot group. When $\delta=0$ the position of the precursor is the same as that of the major flare. In the case of $\delta=1$, the distance between the precursor and the main flare reaches the maximum value because the distance equals to the diameter of the current sunspot group. We assume that $\delta>1$ may only occur in the case of an error as such a value could be due to a wrong active region identification in the flare list or it may be caused by erroneous positioning of the precursor. Hence, all precursor flares with $\delta>1$ have been omitted from the study.

\begin{table}
\centering
\caption{The results of Shapiro-Wilk test for normality based on the distribution $log(\delta)$. The sample size is 100 in all cases. The chosen significance level is 0.05.}
\begin{tabular}{llll}
   X-class & W & p-value & Null hypothesis is  \\
    \hline
    	4th period  &  0.9475  &  0.0005 & rejected. \\
   		3rd period  &  0.9533  &  0.0013 & rejected.\\
    	2nd period &  0.9781  &  0.0951 & not rejected. \\
		1st period  &  0.9851  &  0.3259 & not rejected. 
		
\end{tabular}
\label{sw}
\end{table}

Figure \ref{NFD} shows the normalised frequency distributions of the quantity $\delta$. The left- and right-hand side panels contain the statistics corresponding to M- and X-class flares, respectively. The four rows represent the four 6-hour subintervals of the 24-hour time period. The statistics of precursors corresponding to X-class flares (right panel in Figure \ref{NFD}) shows that there is a dominant peak located roughly at $\delta=0.15$ in each time subinterval. The subplots for the second, third and fourth period show the presence of a co-dominant peak situated at about $\delta=0.55$. Closer in time to the onset of X-class flare, the co-dominant peak gradually vanishes, and it cannot be seen in the 1st period (0-6 hours prior to the X-class flare). The distance between the two peaks can be associated with the distance between the following and leading parts of sunspot groups. This shows that 18-24 hours before an X-class flare, there are usually two main flaring regions at a distance comparable with that of the following and leading parts of sunspot groups. As the time approaches to the onset of the X-class flare, the activity in the co-dominant flaring region decreases, while the number of precursors increases in the dominant region. This implies that in the fourth period a very large part of the active region is involved in the reorganisation of the magnetic structure, after that the reorganisation gradually focuses on a small region of the sunspot group around the location of the coming major flare.

The distributions of precursors of X-class statistics in the 1st, 2nd, 3rd, 4th periods were checked by the Shapiro-Wilk test for normality. The null hypothesis for this test is that the data are normally distributed. If $\delta_{norm}=log(\delta)$ is normally distributed, then the distribution of $\delta$ is called lognormal. The Table \ref{sw} shows the results of the tests. If the p-value is less then the significance level (0.05), then the null hypothesis that the data are normally distributed is rejected. If the p-value is greater than the significance level, then the null hypothesis is not rejected.

In the 1st and 2nd periods the p-value is greater than the significance level, then the null hypothesis cannot be rejected. The distributions could be described by a lognormal distribution thus we call this type of distribution as lognormal-like distribution. In the 3rd and 4th period the test rejected the null hypothesis. We conclude that the distribution is not lognormal in the 3rd and 4th periods with a statistical significance. In the 1st period, the maximum of the distribution corresponds to $\delta=0.09$ (it equals to $11.7\pm 0.8$ Mm derived from the average diameter), which is the most probable distance. $95\%$ of the selected events are closer to the main flare than $\delta=0.46$ ($2\sigma$, equals to $65\pm 4.4$ Mm). 

In the case of M-class flares (left panel in Figure \ref{NFD}), the majority of precursors occur in a small part of the sunspot group around the location of the main flare during the whole day. The lognormal-like distribution seems to be a general property of this class because all the four subplots show this behaviour. This implies that the reorganisation of the magnetic structure leading to an M-class flare is mainly confined to this small part, in contrast to the large-scale restructuring before X-class flares.

\begin{figure}
	\center
	\includegraphics[width=85mm]{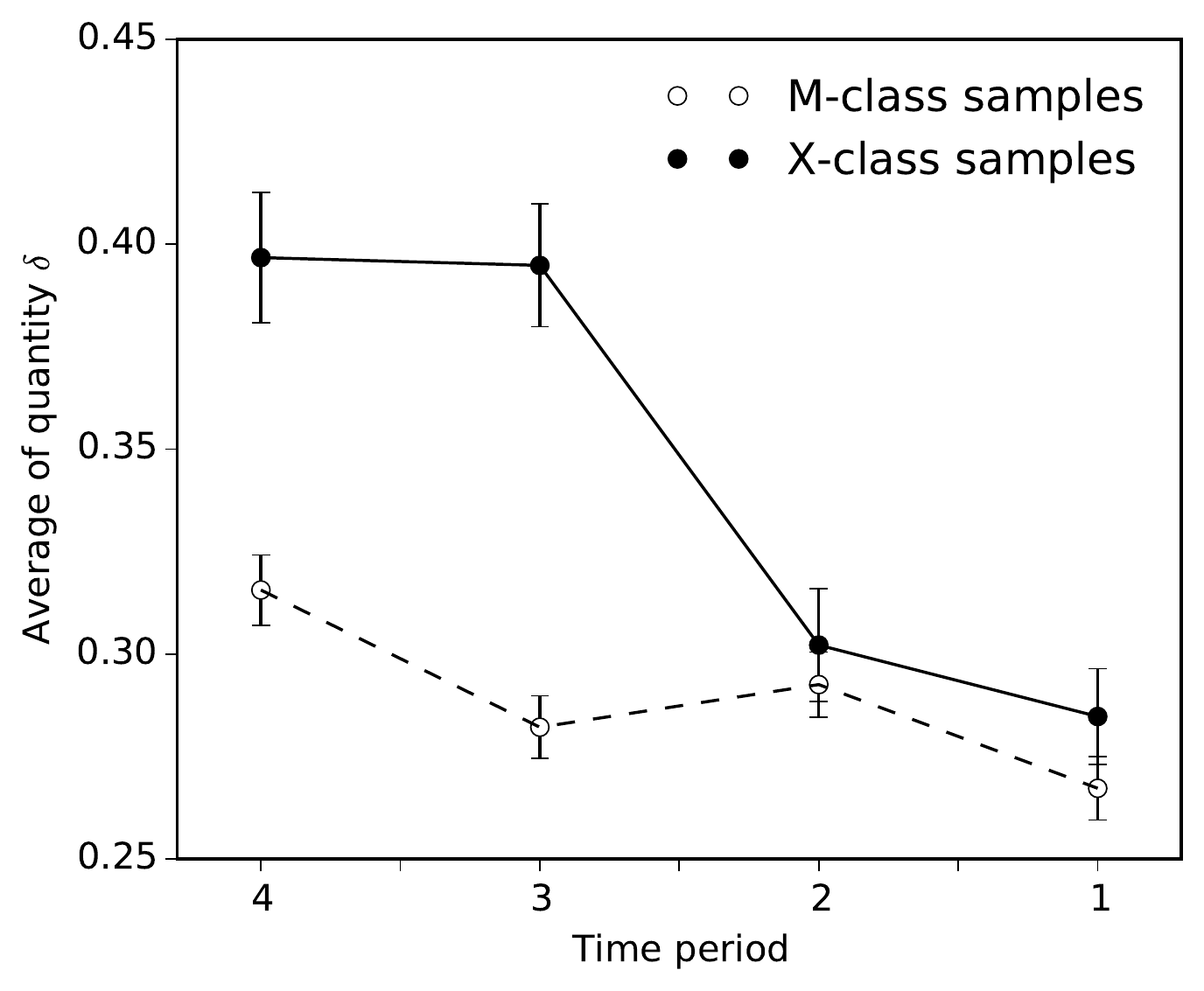}
	\caption{The average of the quantity $\delta$ versus the previously defined time periods. The black circles represent the statistics corresponding to the X-class flares, while the white circles represent the M-class data, both with error bars. The error bars have been estimated by applying Poisson statistics.}
	\label{avg}
\end{figure}

For quantitative comparisons, the average values of the distributions shown in Figure \ref{NFD} can be seen in Figure \ref{avg}. The small average relative distance in the hours before the main flares is in agreement with the results of \cite{Jakimiec} and \cite{Balazs} who concluded that the precursors are often placed in the vicinity of strong flares.

\subsection{Temporal distribution of associated flares}

\begin{figure}
	\center
	\includegraphics[width=85mm]{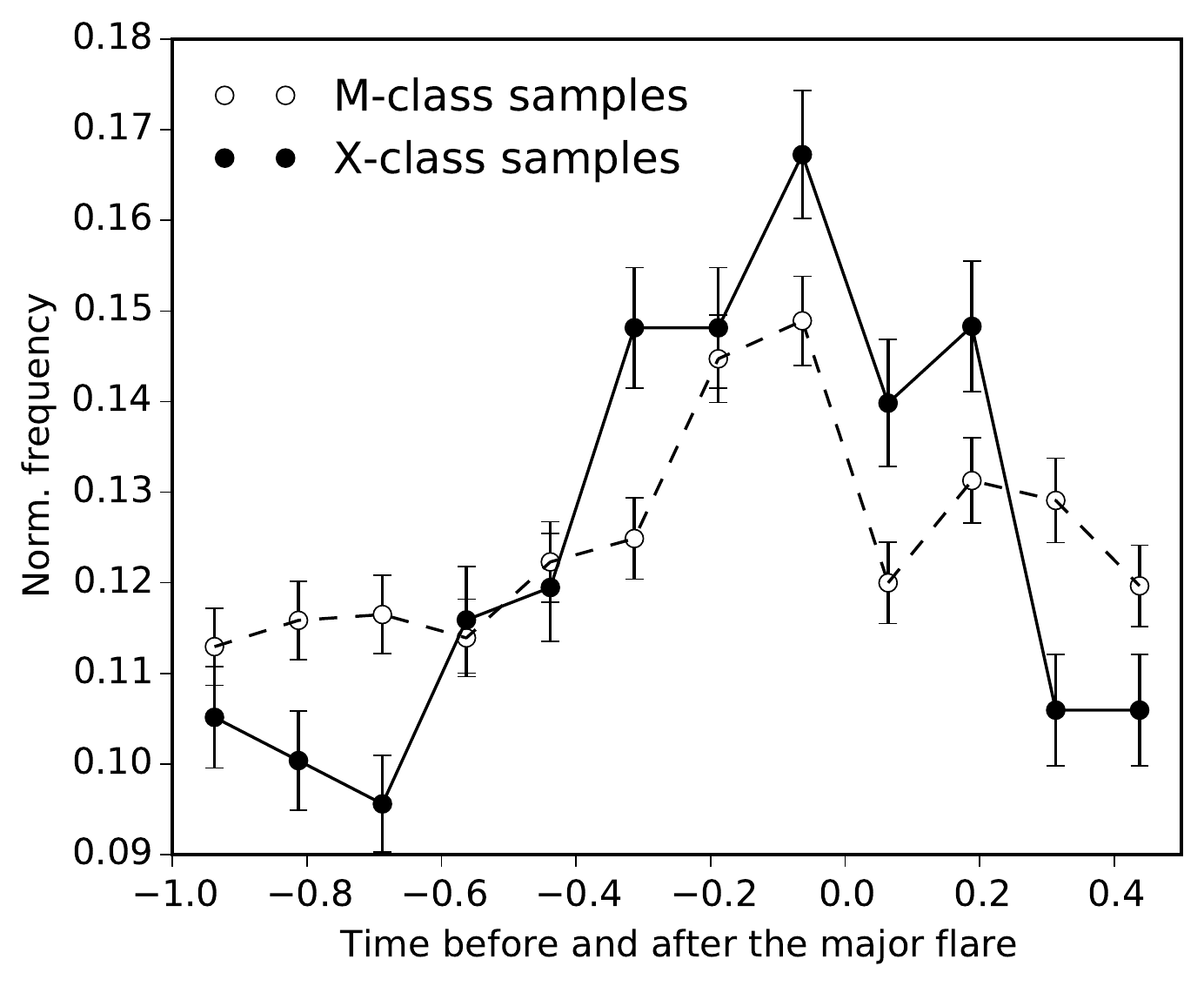}
	\caption{The time variation of the normalised frequency of the number of associated flares. The flares associated with M-class main flares have been indicated by the white circles. The black circles represent the associated  flares occurring in the interval around the X-class flares. The error bars have been estimated by applying Poisson statistics. On the horizontal axis, we plot the time interval under investigation comprising 1 day before the major events and half a day afterward.}
	\label{number_of_flares}
\end{figure}

Here, we investigate the temporal variation of the number of associated flares by applying a superposed epoch analysis.  Figure \ref{number_of_flares} shows the normalised frequency distribution of the number of associated flares in 3-hour bins of the 24-hour period before the main flare and in half a day after the main flare in the case of X-class flares (indicated by black circles) and in the case of M-class flares (indicated by white circles).  Both distributions resemble to bell-shaped curves, but the amplitude is larger in the case of X-class flares. If we assume that the mean of the first three values represents the average basic level of the flaring activity of the given type of flare, then we can estimate that the increased flaring activity lasts about 20-24 hours in both cases.  The main flare occurs at about the middle of this interval but the curves describing the number of associated flares are not symmetrical with respect the time of the main flare. The increase of precursor activity seems to start somewhat earlier in the case of X-class flares than in the case of M-class flares. The characteristic time when the reorganisation of the magnetic structure leading to increased precursor activity starts can be estimated to be about 12-15 hours before X-class flares and about 7-10 hours before M-class flares.  Figure 4  also shows that even after the large energy release, there is still energy stored in the ARs that could initiate follower flares after the main event lasting about half day.  The return of ARs to their initial energetic state is quicker in the case of X-class flares.

The difference is small but it may refer to an effect occurring mainly after large flares when the X-ray background has an increased level for a long time. For the RHESSI flare list, a flare candidate is flagged as a possible solar flare if the ratio of the count rate in the front detectors to total count rate is 3 sigma above its own background level determined using a 60 second running average. \cite{Smith} lists those cases when the usual background estimation is inaccurate.  For example, when the flare is long, the background varies significantly over tens of minutes. For flares with faint hard emission, more precise methods of background estimation are desirable too.  In such or similar cases, the method of RHESSI flare list may result in a smaller number of associated flares during the time of increased X-ray background. This effect may contribute to the small asymmetry of the curve for X-class flares.

\subsection{Validation of our data}

\begin{figure}
\centering
\includegraphics[width=85mm]{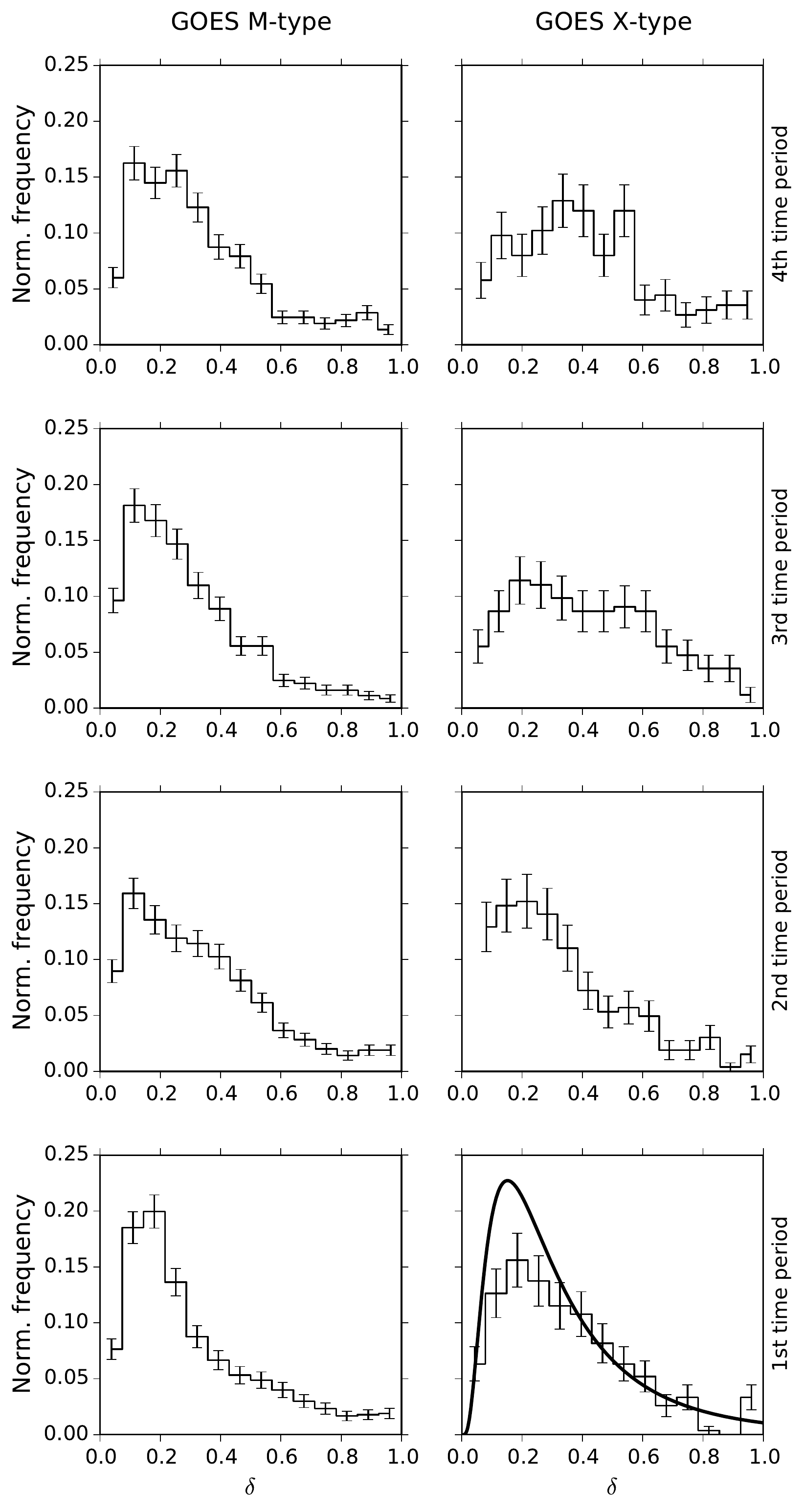}
\caption{Result of the MC simulation. The normalised frequency distribution of quantity $\delta$ in the 24-hour interval prior to the major GOES M- and X-class flares. The influence of the spatial resolution of the RHESSI satellite, the different hight scales and the X-ray albedo effect is modelled.}
\label{mc1}
\end{figure}

Like all studies that deal with measurements involving instruments and data analysis. In this section we perform a Monte-Carlo simulation to reveal the significance of errors and uncertainties that could appear either due to the limitation of the instrument or imaging techniques, or due to undesired physical mechanisms that are able to influence the results (e.g. Compton scattering).

The accuracy of position determination used in Section 2 depends on the employed image processing technique. The RHESSI team introduced several independent methods to estimate the position of the X-ray sources, e.g. Back Projection  \citep{Hurford}, Clean \citep{Hogbom}, Forward Fit \citep{Aschwanden}, Pixon \citep{Pina}, uv-smooth \citep{Massone}, etc. Some of the image reconstruction algorithms can provide source locations within 1 arcsec or better accuracy and they were used in several studies, e.g. \cite{Liu, Kontar2008}. However, these methods may have several disadvantages such as photometric errors, extremely slow running time or require subjective smoothing \citep{Hurford}. For these reasons, these algorithms are usually used in particular case studies; in contrast our paper considers an enormous amount of data. The RHESSI flare list is a very detailed X-ray survey, containing sufficient amount of data for a statistical analysis. The position of flares are indicated with precision of one arcsecond in the list but these data are based on 128x128 back-projection maps using 16-arcsecond pixels created by Back Projection algorithm. Thus, the precision of the position data is approximately 16 arcseconds (J. McTiernan 2016, personal communication). The spatial resolution might seem to be a bit low but the investigated sunspot groups (which are able the produce a very energetic solar flare) are one magnitude larger than the 16 arcsecond resolution. Therefore the estimated error of the position in RHESSI flare list is less than 10$\%$ of the investigated sunspot groups.
 
The other effect which is able to influence the observed sizes and also positions of X-ray sources is the Compton scattering or X-ray albedo. \cite{Kontar2010} used a Monte-Carlo simulation to investigate the significance of the Compton backscattering and they showed that this effect can influence the accuracy of the positions and source sizes of X-ray sources and, therefore, should always be taken into account. They also stated that due to these effects, there is a radially directed displacement towards the disk centre. This shift is also able to affect the position of the sources when we estimate the heliographical coordinates, therefore, this could be significant in our study. The largest displacement can be observed for events in the $30 - 50$ keV range at 60 degrees from the central meridian. The shift in centroid position in this energy range is $0.1 - 0.5$ arcsec ($0.09 - 0.39$ Mm). The displacement of HXR source position is energy dependent.

Finally, the reconnection of magnetic fields resulting in a flare does not occur on the surface of the Sun. It is a crucial point to estimate the height of flares. \cite{Takakura} measured a hard X-ray source height of h = $7.0 \pm 3.5$ Mm above the neutral line seen in H$_{\alpha}$ in the 20 - 40 keV range with Hinotori. \cite{Matsushita} measured the statistical displacement of hard X-ray source centroids to reported H$_{\alpha}$ flare positions and found heights of  $9.7 \pm 2.0$ Mm for flares in the 14 - 23 keV energy band. The distribution of HXR (4 - 10 keV) source heights was found to be well fitted by an exponential distribution with a scale height of $6.1 \pm 0.3$ Mm. \cite{Christe} concluded that the minimum height due to partially occulted sources is $5.7 \pm 0.3$ Mm.

In order to describe the uncertainties due to position determination, the height of the event and the Compton scattering, we assume three PDFs describing the appearance of errors in our position estimations. The first point we are addressing is the uncertainty caused by the spatial resolution of the RHESSI flare list. We do not know where is the exact position of the solar flare inside a pixel, which means it could be anywhere in a 16$\times$16 arcsecond$^2$ sized box. For that reason, the PDF is a uniform distribution; having an equal chance of being located anywhere inside of the box. The second uncertainty is introduced by the height of the reconnection. Based on the study by \cite{Christe}, the source heights was found to be $6.1 \pm$ 0.3 Mm, so we take 6.1 Mm as a most probably value and  $\pm$ 0.3 Mm as a uncertainty. These suggest a normal distribution. Finally the shift of the X-ray source in centroid position due to Compton scattering is $0.1 - 0.5$ arcsecond. Since we do not know the exact shape of the distribution between these two values, we assume a step function.

 Figure \ref{mc1} shows the result of the Monte - Carlo simulation. The distributions based on the GOES M-type statistics shows very similar properties as the results shown in Figure \ref{NFD}. Most of the microflares concentrate near the major flare. The other group based on GOES X-type flares shows slightly different spatial properties. In the 4th and 3rd time interval the double-peak distributions are not so obvious. However, the meaning is the same: the subplots for the second, third and fourth period show the presence of a co-dominant peak (plateau) situated at about $\delta = 0.55$. The Monte-Carlo simulation based on the three PDFs reveal that distribution of the normal frequency of appearance of flares with respect to the quantity $\delta$ is similar to the distribution obtained in Figure \ref{NFD}, confirming the robustness of our results.

\begin{figure}
\centering
\includegraphics[width=85mm]{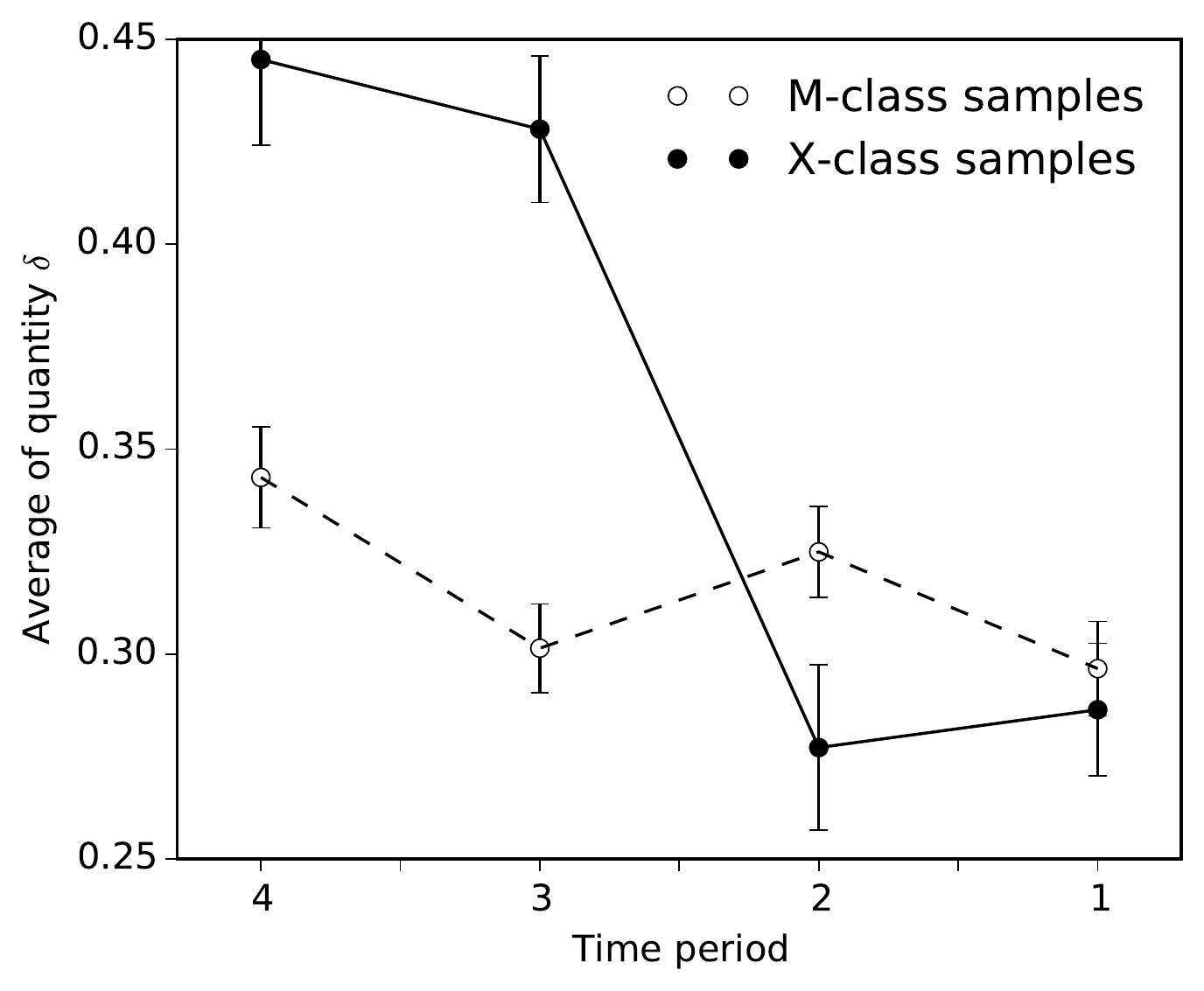}
\caption{The average of the quantity $\delta$ versus the previously defined time periods. The black circles represent the statistics corresponding to the X-class flares, while the white circles represent the M-class data, both with error bars. We omitted all those candidates that erupted 35 degrees further from the central meridian.}
\label{mc2}
\end{figure}

We also checked whether the centre-to-limb variation of the error of flare position (because of the increasing projection effect and decreasing spatial resolution) affects our results.  We repeated the same analysis as in Section 3 by using the selection criterion $abs(LCM) < 35^{\circ}$. In this case, the sample of data is smaller but the accuracy of position data is larger. The results of the statistics are plotted in Figure \ref{mc2}, which shows similar trends for the two classes of flares as Figure \ref{avg}. The differences between the curves of the two figures are about a few hundredths of $\delta$. This means that the effect of the centre-to-limb variation of the error of flare position on the result is small if we use the selection criterion of $abs(LCM) < 70^{\circ}$. We conclude that, the reliability of our result has been validated.

\section{Summary and Conclusion}

We investigated statistically the spatial and temporal distributions of precursor and follower flares in the 24-hour time interval before and after major flares. The study is based on a very large number of events (364 main flares, 4152 precursor flares and 1857 follower flares), trying to establish whether there is any statistically significant detectable character in the evolution of precursor flare activity. Our study is a major step forward in field of flare diagnostics as it addresses the question of spatial and temporal occurrence of flares and the temporal evolution of preceding flares 24 hour prior to the eruption of an X- or M-type flare. Furthermore, our investigation provides information on such a segment of the observational phase space that has not been investigated so far. We determined parameters that could help in describing quantitatively the changes occurring in ARs or can be suitable for observational constraints for theoretical models.

\newpage

The spatial distribution was characterised by the normalised frequency distribution of the quantity $\delta$ (the distance between the major flare detected by GOES satellite and its precursor flare observed by RHESSI normalised by the sunspot group diameter) in four 6-hour intervals before the major event. The precursors of X-class flares initially follow a double peaked spatial distribution but the co-dominant peak gradually vanishes and the shape of the distribution changes to a lognormal-like distribution roughly 6 hours prior to the event. The quantity $\delta$ belonging to the precursors of M-class flares show lognormal-like distribution in each 6-hour subinterval. These statistical results support the idea that the reorganisation of the magnetic structure of an AR has a larger spatial extent before X-class flares than M-class flares.

Our result shows that the most frequent sites of the precursors in the AR are within a distance of about 0.1 diameter of sunspot group from the site of the forthcoming major flare. This also means that the precursors may indicate the position of the main flare with about this accuracy several hours before it occurs.  The average of the relative distance, $\delta$, of precursors from the site of the main flare is about 0.3 diameter during the half a day before the main flare.

The temporal variation of the number of associated flares resembles to a bell-shaped curve for both X-class and M-class main flares. The increased flaring activity also last about 20-24 hours in both cases.  The increasing and decreasing periods of the curves are not equal. In the case of X-class main flares, the precursor activity shows an earlier starting and larger increase than in the case of M-class flares. The characteristic time when the reorganisation of the magnetic structure leading to increased precursor activity starts can be estimated to be about 12-15 hours before X-class flares and about 7-10 hours before M-class flares. The main flare occurs at about or after the maximum of the number of associated flares. This may show that the build-up of energy is much more effective than the release of energy because of precursors flares. After the main flare, the number of follower flares decreases more quickly in the case of X-class flares. This may show that the X-class flare decreases the energy stored in the AR in such an extent that the return of ARs to their initial energetic state can be quicker than in the case of M-class flares. The results imply that all the main flare and the associated flares are parts of a dominant underlying process of energy build-up and release lasting about 20-24 hours in average.  

The robustness of our investigation was tested employing worst-case scenario error sources in the determination of the position of flares and our simulations show that even if these sources are taken into account the patterns recovered for the spatial distribution of precursor flares still stand. 

The novelty of our research resides in the fact that the employed statistical investigations revealed a number of quantitative spatial and temporal characteristics of the as-yet-unknown dominant underlying process (or series of processes) that drives the set of successive flares containing the major flare and the associated flares in the period about one day around the major flare. The statistically derived values may help in determining the key mechanisms playing role in the studied interval from a few hours to about one day. In order to better explain our findings (and their implications), we would also need a much deeper understanding of underlying physical mechanisms and the reorganisation of the structure of the coronal magnetic field before major flares, for which numerical field reconstruction models are needed. 

\section*{Acknowledgements}

We would like to thank L. Fletcher and J. McTiernan for all useful suggestions. N. Gyenge (NG) would like to acknowledge the hospitality of the SP$^2$RC group members while visiting The University of Sheffield. NG also thanks the ERASMUS Student Exchange Program for the provided support. The GOES "X-ray Flare" dataset was prepared by and made available through the NOAA National Geophysical Data Center (NGDC). This research has made use of SunPy, an open-source and free community-developed solar data analysis package written in Python \citep{SunPy}. I. Ballai acknowledges the support by the Leverhulme Trust (IN-2014-016) and ANCNS Romania (STAR, 72/2013). The research leading to these results has received funding from the European Community's Seventh Framework Programme (FP7/2007 - 2013) under the eHEROES grant agreement  (project No. 284461).

\end{document}